\documentclass[groupedaddress,aps,pra,superscriptaddress,showpacs,twocolumn,prl]{revtex4}%
\usepackage{epsfig,dsfont,amssymb,amsmath,amsthm,amsfonts,amsbsy,mathrsfs}
\usepackage{graphicx, color}
\usepackage{epstopdf}
\usepackage{mathdots}
\usepackage{float}
\usepackage{xcolor}
\begin{document}

\title{Polygamy relation of quantum correlations with equality}

\author{Zhi-Xiang Jin}
\thanks{Corresponding author: jzxjinzhixiang@126.com}
\affiliation{School of Computer Science and Technology, Dongguan University of Technology, Dongguan, 523808, China}
\author{Bing Yu}
\thanks{Corresponding author: mathyu590@163.com}
\affiliation{School of Mathematics and Systems Science, Guangdong Polytechnic Normal University, Guangzhou,510665,China}
\author{Xue-Na Zhu}
\thanks{Corresponding author: jing\_feng1986@126.com}
\affiliation{School of Mathematics and Statistics Science, Ludong University, Yantai 264025, China}

\author{Shao-Ming Fei}
\thanks{Corresponding author: feishm@cnu.edu.cn}
\affiliation{Max-Planck-Institute for Mathematics in the Sciences, Leipzig 04103, Germany}
\affiliation{School of Mathematical Sciences, Capital Normal University,  Beijing 100048,  China}

\author{Cong-Feng Qiao}
\thanks{Corresponding author: qiaocf@ucas.ac.cn}
\affiliation{School of Physics, University of Chinese Academy of Sciences, Yuquan Road 19A, Beijing 100049, China}
\affiliation{Key Laboratory of Vacuum Physics, Chinese Academy of Sciences, China\\ \vspace{7pt}}

\begin{abstract}
We provide a generalized definition of polygamy relations for any quantum correlation measures. Instead of the usual polygamy inequality, a polygamy relation with equality is given by introducing the polygamy weight. From the polygamy relation with equality, we present polygamy inequalities satisfied by the $\beta$th $(\beta>0)$ power of the quantum correlation measures. Taking concurrence of assistance as an example, we further illustrate the significance and advantages of these relations. We also obtain a polygamy relation with equality by considering the one-to-group entanglements for any quantum entanglement measures that do not satisfy the polygamy relations. We demonstrate that such relations for tripartite states can be generalized to multipartite systems.
\end{abstract}
\maketitle

\section{Introduction}
As one of the essential resources in quantum information processing, quantum correlation is widely used in quantum tasks, such as quantum teleportation \cite{wootter}, quantum key distribution \cite{vss,MP}, and entanglement swapping \cite{bd,maa}. The fundamental difference between quantum correlation and classical correlation lies in the shareability of the resources. With classical correlations, the resources can be freely shared by many different individuals, while for quantum correlations, the resources cannot be shared among individuals freely. Therefore, it is particularly important to study the quantum correlation shareability or distribution among multipartite systems and their implications in quantum information processing and quantum computing \cite{jzx,ZXN,ckw,jinzx,MK,jin1}.

Quantum entanglement, one of the most important quantum correlations, plays a crucial role in quantum communication and quantum information processing \cite{MAN,RPMK,FMA,KSS,HPB,HPBO,JIV,CYSG}. The restriction on entanglement shareability among multiparty systems is called the monogamy of entanglement; that is, with respect to a given entanglement measure, if two subsystems are more entangled, they would be less entangled with the rest subsystems. The monogamy of entanglement is a crucial property that guarantees the quantum key distribution security since the limited shareability restricts the amount of information that an eavesdropper could potentially obtain by the secret key extraction.
It also plays an imperative role in many other fields of physics, such as quantum cryptography \cite{jmr,ml}, phase detection \cite{max,kjm,gsm}, condensed matter physics \cite{ma,aj}, and even black-hole physics \cite{ls,sj}.

Conversely, the entanglement of assistance, a dual concept to bipartite entanglement measures, has been shown to have a dually monogamous (polygamous) property in multipartite quantum systems. Polygamy inequality was first obtained in terms of the tangle $E_a$ of assistance \cite{gour} among three-qubit systems. For a three-qubit state $\rho_{ABC}$,
\begin{eqnarray}\label{m1}
E_a(\rho_{A|BC})\leq E_a(\rho_{AB})+E_a(\rho_{AC}),
\end{eqnarray}
where $\rho_{AB}=\mathrm{Tr}_C(\rho_{ABC})$ and $\rho_{AC}=\mathrm{Tr}_B(\rho_{ABC})$.
Intuitively, Eq. (\ref{m1}) indicates that the entanglements of assistance between $A$ and $BC$ cannot exceed the sum of the individual pairwise entanglements of assistance between $A$ and each of the remaining parties $B$ or $C$. In fact, monogamy relations provide an upper bound for bipartite shareability of quantum correlations in a multipartite system, while polygamy relations set a lower bound for the distribution of bipartite quantum correlations. Variations of the polygamy inequality and the generalizations to $N$-partite systems have been established for several quantum correlations \cite{jzx1,jin2,kim,guoy}. Nevertheless, to some extent, the inequality (\ref{m1}) captures the spirit of polygamy as a distinctive property of entanglement of assistance since its validity is not universal but rather depends on the detailed quantum states, the kind of correlations, and the measures used. For example, the GHZ-class states do not satisfy the inequality (\ref{m1}) for any quantum entanglement measures.

In this paper, we study the general polygamy relations of arbitrary quantum correlations. Given a measure $Q$ of general quantum correlation, we classify it as polygamous if there exists a nontrivial continuous function $g$ such that the generalized polygamy relation,
\begin{eqnarray}\label{m2}
Q(\rho_{A|BC})\leq g(Q(\rho_{AB}), Q(\rho_{AC})),
\end{eqnarray}
is satisfied for any state $\rho_{ABC}$, where $g$ is a continuous function of variables $Q(\rho_{AB})$ and $Q(\rho_{AC})$.
For convenience, we denote $Q(\rho_{A|BC})=Q_{A|BC}$ and $Q(\rho_{AB})=Q_{AB}$.
For a particular choice of the function $g(x,y)=x+y$, we recover the polygamy inequality (\ref{m1}) from (\ref{m2}). As $Q$ is a measure of quantum correlation, it is nonincreasing under partial trace. Thus, we have $Q_{A|BC}\geq \max\{Q_{AB}, Q_{AC}\}$ for any states. Specifically, the quantum correlation distribution is confined to a region smaller than a square with a side length $Q_{A|BC}$. For any state $\rho_{ABC}$, if there exists a nontrivial function $g$ such that the generalized polygamy equality $Q_{A|BC}= g(Q_{AB}, Q_{AC})$ is satisfied, we call the quantum correlation measure $Q$ polygamous.

\section{Polygamy Relation with Equality and its Properties}
Consider the function $g(x,y)=x+y$ as a rubber band. For two fixed endpoints $(Q_{A|BC},0)$ and $(0, Q_{A|BC})$, one obtains different types of functions by moving the point $(\frac{Q_{A|BC}}{2},\frac{Q_{A|BC}}{2})$ to the point $(Q_{A|BC},Q_{A|BC})$ or to the origin $(0,0)$, as shown in Fig. 1. For any $0< k\leq Q_{A|BC}$, $(k,k)$ is a point on the dotted line in Fig. 1. We have the following trade-off between the values of $Q_{AB}$ and $Q_{AC}$,
\begin{equation}\label{m3}
    \begin{cases}
    \frac{Q_{A|BC}-k}{k}Q_{AB}+Q_{AC} = Q_{A|BC}, ~~ Q_{AB}\leq Q_{AC}\\[2mm]
       Q_{AB}+ \frac{Q_{A|BC}-k}{k}Q_{AC} = Q_{A|BC},~~Q_{AC}\leq Q_{AB},
     \end{cases}
\end{equation}
where ``k" is different for different lines in Fig. 1, but the same with the symmetric lines on both sides of the dashed line.
In fact, we only need to consider the range of $0<k\leq \frac{Q_{A|BC}}{2}$ because the original polygamy inequality is always satisfied for $\frac{Q_{A|BC}}{2}< k\leq Q_{A|BC}$ (blue region in Fig. 1). The equation with respect to the diagonal solid line in the square is given by $Q_{AB}+Q_{AC}= Q_{A|BC}$, and the triangle area above the diagonal solid line (blue region) satisfies $Q_{AB}+Q_{AC}> Q_{A|BC}$. On the contrary, the triangle area below the diagonal solid line (orange, yellow and white regions) satisfies the inequality $Q_{AB}+Q_{AC}< Q_{A|BC}$. The distribution of quantum correlation for all quantum states should be checked to verify whether a quantum correlation measure $Q$ is polygamous. If the quantum correlation for all states is distributed on the upper triangle area, it means that $Q$ satisfies the polygamy relations.

Due to the importance of polygamy relations, it is also interesting to ask whether a quantum correlation measure $Q$ with some quantum correlation distribution in the lower triangular region can be superactivated to be polygamous. If $\rho$ does not satisfy the polygamous relations with respect to some quantum correlation measure $Q$, can some of these quantum correlation measures be polygamous, i.e., is there a hidden polygamy relation?
Thus, it is interesting to study a quantum correlation measure $Q$ when its quantum correlation is distributed in the lower triangular region and determine ways to characterize its polygamy relation.
Inspired by Ref. \cite{jin}, we propose a parameterized polygamy relation of quantum correlation with equality, which is more powerful than the original ones. From Eq. (\ref{m3}), we can define a polygamy relation with equality as follows.

{\bf Definition 1.} Let $Q$ be a measure of quantum correlation. $Q$ is said to be polygamous if for any state $\rho_{ABC}$,
\begin{eqnarray}\label{m4}
Q_{A|BC}= \gamma Q_{AB}+Q_{AC}
\end{eqnarray}
for some $\gamma$ $(\gamma\geq0)$ with $Q_{AB}\leq Q_{AC}$. We call $\gamma$ the polygamy weight with respect to the quantum correlation measure $Q$.

Eq. (\ref{m4}) yields a generalized polygamy relation without inequality. For different states, one can always obtain different $\gamma$. Thus, $\gamma$ in Eq. (4), in fact, is the biggest constant taken over all states that saturate the above equality. The polygamy weight $\gamma$ defined in Eq. (4) establishes the connections among $Q_{A|BC}$, $Q_{AB}$ and $Q_{AC}$ for a tripartite state $\rho_{ABC}$. If $0\leq \gamma\leq1$, then the polygamy inequality (\ref{m1}) is obviously true from (\ref{m4}), and the corresponding quantum correlation distribution is confined to the blue region shown in Fig. 1. The case of $\gamma>1$ is beyond the original polygamy inequality and the corresponding regions of the quantum correlation distribution are in the orange, yellow and white regions in Fig. 1.
When $\gamma\to \infty$, we have $Q_{A|BC}> Q_{AC}\geq Q_{AB}=0$ according to the definition (\ref{m4}). In this situation, we say that the quantum correlation measure $Q$ is non-polygamous. The corresponding quantum correlation distribution is located at the coordinate axis in Fig. 1; that is, when $\gamma\to \infty$, $Q$ is not likely to be polygamous.
By contrast, $\gamma\to 1$ implies that $Q$ is more likely to be polygamous.
In Ref. \cite{jin}, the authors present a monogamy weight $\mu$, which can be
viewed as a dual parameter to the polygamy weight $\gamma$, where $\mu\to 1$ implies
that the entanglement measure $E$ is more likely to be monogamous, while $\mu\to 0$ means that $E$ is not likely to be monogamous. If an entanglement measure $E$ is not only polygamous but also monogamous, then it must satisfy $\mu=\gamma=1$. This happens for some classes of states and entanglement measures, such as $W$-class states with the entanglement measure concurrence or negativity.
\begin{figure}
\centering
\includegraphics[width=9cm]{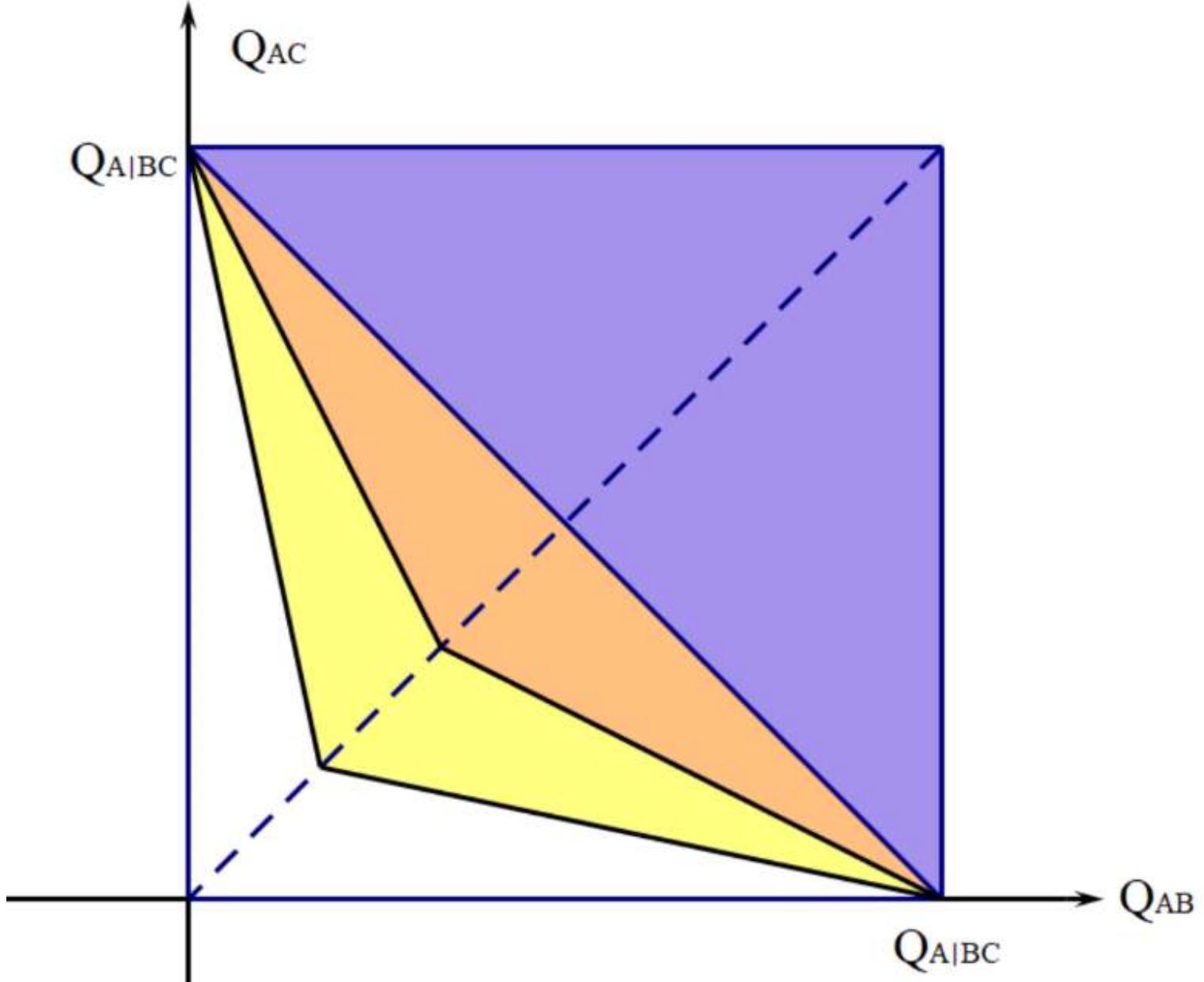}\\
\caption{For any tripartite state $\rho_{ABC}$ and quantum correlation measure $Q$, one gets the inequality (\ref{m1}) for $g(x,y)=x+y$, which holds with the range of values of $Q_{AB}$ and $Q_{AC}$ given by the blue triangular. In the blue region, the equality (\ref{m4}) also holds for $0\leq \gamma\leq 1$. Inequality is no longer satisfied in the red, yellow, and white regions. However, the relation (\ref{m4}) holds for $\gamma>1$: the orange region matches $1< \gamma \leq2$, the yellow region matches $2< \gamma \leq 3$, and the white region matches $\gamma>3$. In other words, any quantum correlation measure $Q$ is polygamous in the sense of (\ref{m4}) if the quantum correlation distribution is confined to a region strictly smaller than the square with side length $Q_{A|BC}$.}\label{2}
\end{figure}

Thus, the parameter $\gamma$ has an operational interpretation of the ability to be polygamous for a quantum correlation measure $Q$. Given two quantum correlation measures $Q'$ and $\tilde{Q}$ with polygamy weights $\gamma_1$ and $\gamma_2$, respectively. We say that $Q'$ has a higher polygamy score than $\tilde{Q}$ if $\gamma_1\geq \gamma_2$. In contrast to the monogamy score proportional to its monogamy ability in Ref. \cite{jin}, the polygamy ability is inversely proportional to the magnitude of its weight score. That is, $\gamma_1\geq \gamma_2$ leads to ${Q'}\preceq \tilde{Q}$, where ${Q'}\preceq \tilde{Q}$ stands for that $\tilde{Q}$ has stronger ability than $Q'$ to be polygamous. Thus, $\gamma$ characterizes the polygamy property of a given quantum correlation measure $Q$.
We have the following relation for the $\beta$th power $Q^\beta$ ($\beta>0$) of $Q$ (see the proof in the Appendix).

{\bf Theorem 1} A quantum correlation measure $Q$ is polygamous according to definition (\ref{m4}) if and only if there exits $\beta > 0$ such that
\begin{eqnarray}\label{Theorem1}
Q^\beta_{A|BC}\leq Q^\beta_{AB}+Q^\beta_{AC}
\end{eqnarray}
for any state $\rho_{ABC}$.

{\it Remark.}  Theorem 1 shows that the polygamy weight $\gamma$ in Eq. (\ref{m4}) has a one-to-one correspondence to the polygamy power in inequality (\ref{Theorem1}) for a given quantum correlation measure $Q$. In \cite{jinzx}, the authors show that there exists a real number $p$ such that $Q^y$ $(0\leq y\leq p)$ is polygamous. That is to say, the polygamy weight $\gamma$ corresponds one-to-one to $p$ in \cite{jinzx}. Combining Eq. (\ref{m4}) and (\ref{Theorem1}), one gets $\gamma= (1+K^\beta)^\frac{1}{\beta}-K$ with $K=\frac{Q_{AC}}{Q_{AB}}>1$ if $\beta$ saturates the inequality (\ref{Theorem1}), which implies that the polygamy weight $\gamma$ and the real number $p$ in \cite{jinzx} are inversely proportional (see Fig. 2). In \cite{jin} the authors give a monogamy relation of entanglement $E$, $E_{A|BC}= \mu E_{AB}+E_{AC}$ with $E_{AB}\leq E_{AC}$. At first glance, it looks the same as the one in our Definition 1; however, the ranges of monogamy weight $\mu$ and polygamy weight $\gamma$ are completely different. In \cite{jin}, the monogamy weight takes the value $0<\mu\leq 1$, since $\mu>1$ is obviously true for CKW inequality \cite{ckw}. By contrast, the range of the polygamy weight $\gamma$ in this paper is $\gamma>1$ since it is obviously true for $0\leq \gamma\leq 1$. This may explain the difference between monogamy and polygamy relations based on polygamy (monogamy) weight $\gamma$ $(\mu)$ from another perspective.
\begin{figure}
\centering
\includegraphics[width=8.5cm]{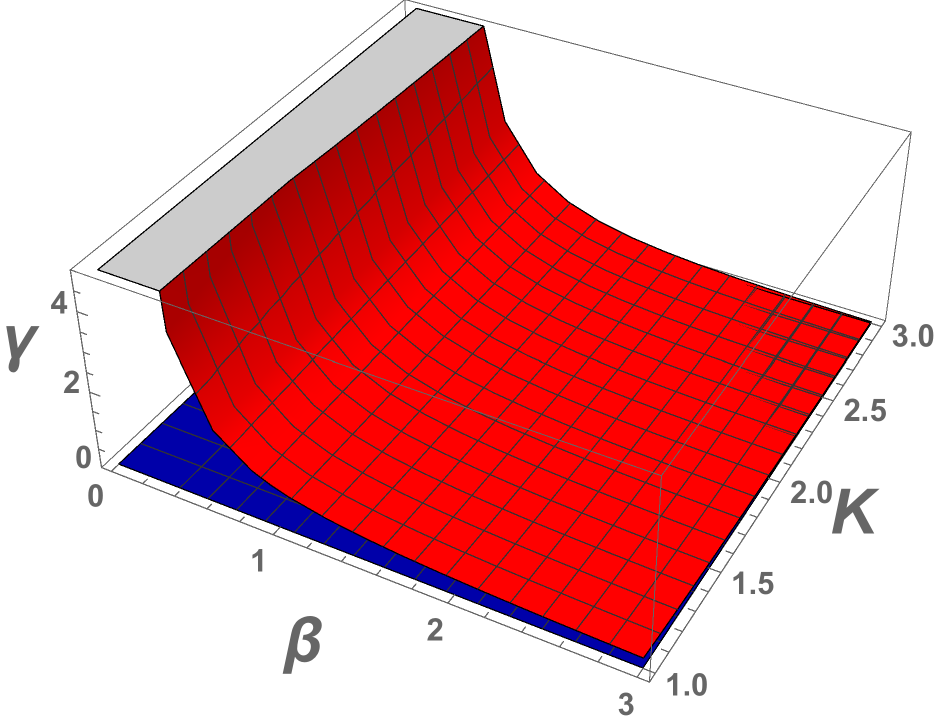}\\
\caption{One-to-one mapping between the polygamy weight $\gamma$ in Eq. (\ref{m4}) and the polygamy power $p$ in \cite{jinzx} for a given quantum correlation measure $Q$.}\label{2}
\end{figure}

The polygamy relation defined in Eq. (\ref{m4}) can be generalized to multipartite systems. For any $N$-partite state $\rho_{AB_1B_2...B_{N-1}}$, we obtain the following result if $Q$ satisfies Eq. (\ref{m4}) for any tripartite state (see the proof in Appendix).

{\bf Theorem 2}.  For any $N$-partite state $\rho_{AB_1\cdots B_{N-1}}$, generally assume that ${Q_{AB_i}}\geq {Q_{A|B_{i+1}\cdots B_{N-1}}}$ for $i=1, 2, \cdots, m$, and
${Q_{AB_j}}\leq {Q_{A|B_{j+1}\cdots B_{N-1}}}$ for $j=m+1,\cdots,N-2$, $\forall$ $1\leq m\leq N-3$, $N\geq 4$. If $Q$ satisfies relation (\ref{m4}) for tripartite states, then
\begin{eqnarray*}\label{th1}
&&Q_{A|B_1B_2\cdots B_{N-1}}\nonumber \\
&&\leq Q_{AB_1}+\Gamma_1Q_{AB_2}+\cdots+\Gamma_{m-1}Q_{AB_m}\nonumber\\
&&+\Gamma_m(\gamma_{m+1}Q_{AB_{m+1}}+\cdots+\gamma_{N-2}Q_{AB_{N-2}}+E_{AB_{N-1}}),
\end{eqnarray*}
where $\Gamma_k=\Pi_{i=1}^k\gamma_i$, $k=1,2,\cdots,N-2$, and $\gamma_i$ denotes the polygamy weight of the $(N+1-i)$-partite state $\rho_{AB_1\cdots B_{N-i}}$.

In Theorem 2 we have assumed that some ${Q_{AB_i}}\geq {Q_{A|B_{i+1}\cdots B_{N-1}}}$ and some
${Q_{AB_j}}\leq {Q_{A|B_{j+1}\cdots B_{N-1}}}$ for the $N$-partite state $\rho_{AB_1\cdots N_{N-1}}$. If all ${Q_{AB_i}}\geq {Q_{A|B_{i+1}\cdots B_{N-1}}}$ for $i=1, 2, \cdots, N-2$, then we have $Q_{A|B_1\cdots B_{N-1}}=Q_{AB_1}+\Gamma_1Q_{AB_2}+\cdots+\Gamma_{N-2}Q_{AB_{N-1}}$.

In the following, as an application, we consider the concurrence of assistance to illustrate the advantages of (\ref{m4}) and the calculation of the polygamy weight $\gamma$.
For a bipartite state $\rho_{AB}$, the concurrence of assistance is defined by
$C_a(\rho_{AB})=\max\limits_{\{p_i,|\psi_i\rangle_{AB}\}}\sum_ip_iC(|\psi_i\rangle_{AB})$,
where the maximum is taken over all possible pure state decompositions of $\rho_{AB}=\sum\limits_{i}p_i|\psi_i\rangle_{AB}\langle\psi_i|.$
For any pure states $\rho_{AB}=|\psi\rangle_{AB}\langle\psi|$, one has $C_a(\rho_{AB})=C(|\psi\rangle_{AB})$, where $C(|\psi\rangle_{AB})$ is the concurrence of the state $|\psi\rangle_{AB}$, $C(|\psi\rangle_{AB})=\sqrt{{2\left[1-\mathrm{Tr}(\rho_A^2)\right]}}$, and $\rho_A=\mathrm{Tr}_B(|\psi\rangle_{AB}\langle\psi|)$.
For convenience, we denote $C_a(\rho_{AB})={C_a}_{AB}$ and
${C_a}(\rho_{A|BC}) ={C_a}_{A|BC}$.

Let us consider the three-qubit state $\rho=|\psi\rangle\langle\psi|$ in the generalized Schmidt decomposition form,
\begin{eqnarray}\label{Ex}
|\psi\rangle&=&\lambda_0|000\rangle+\lambda_1e^{i{\varphi}}|100\rangle+\lambda_2|101\rangle \nonumber\\
&&+\lambda_3|110\rangle+\lambda_4|111\rangle,
\end{eqnarray}
where $\lambda_i\geq0$, $i=0,1,2,3,4$ and $\sum\limits_{i=0}\limits^4\lambda_i^2=1.$ We have
${C_a}_{A|BC}=2\lambda_0\sqrt{{\lambda_2^2+\lambda_3^2+\lambda_4^2}},$
${C_a}_{AB}=2\lambda_0\sqrt{{\lambda_2^2+\lambda_4^2}}$ and ${C_a}_{AC}=2\lambda_0\sqrt{{\lambda_3^2+\lambda_4^2}}$.
According to the polygamy relation (\ref{m4}), we have
\begin{eqnarray}\label{ex1}
\gamma=\sqrt{1+\frac{\lambda_3^2}{\lambda_2^2+\lambda_4^2}}-\sqrt{\frac{\lambda_3^2+\lambda_4^2}{\lambda_2^2+\lambda_4^2}}
\end{eqnarray}
with $\lambda_2\leq \lambda_3$. Let $g(x,y)=\sqrt{1+\frac{y^2}{1+x^2}}
-\sqrt{\frac{1+y^2}{1+x^2}}$, with $x=\frac{\lambda_2}{\lambda_4}$
and $y=\frac{\lambda_3}{\lambda_4}$. We obtain $\gamma\leq\sqrt{2}-1$,
with the equality saturated when $x\to \infty,~ y\to \infty$
(i.e., $\lambda_4=0$, $\lambda_2=\lambda_3\ne0$) (see Fig 3).
In other words, the W-type states ($\lambda_4=0$ in (\ref{Ex})) saturate the maximal value
of the polygamy weight for concurrence of assistance in Eq. (\ref{ex1}), $\gamma_{C_a}=\sqrt{2}-1<1$. The corresponding quantum correlation distribution is
confined to the blue region in Figure 1. From the conclusion in \cite{zhuxn,jzx2},
the W-type states just satisfy the equality in (\ref{Theorem1}), and the corresponding polygamy power is $\beta_{C_a}=2$.
\begin{figure}
\centering
\includegraphics[width=8cm]{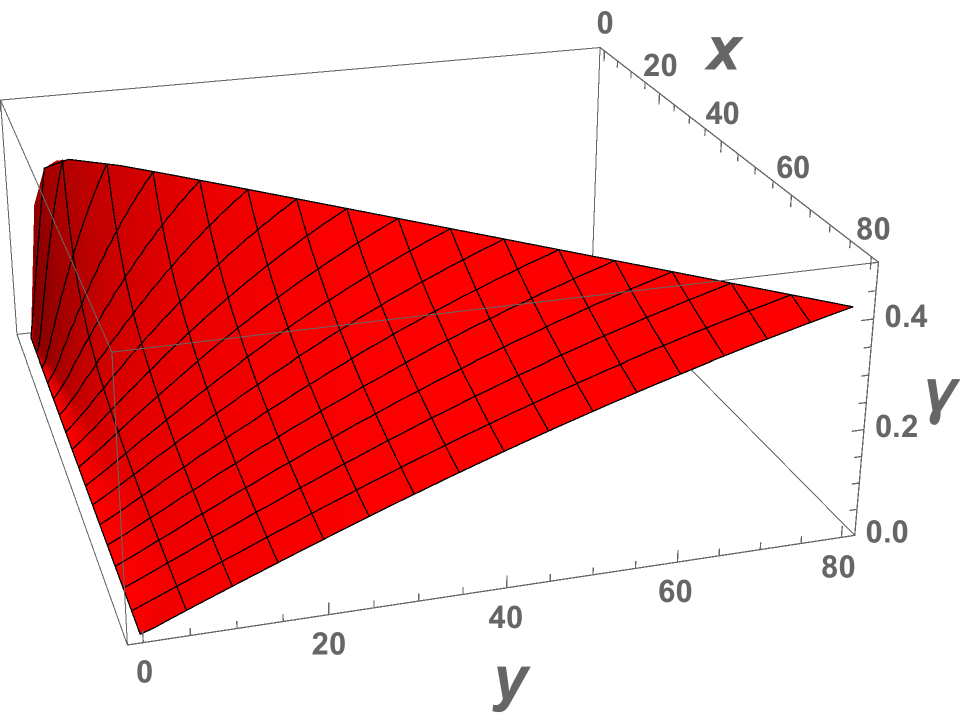}\\
\caption{Polygamy weight $\gamma$ of three-qubit pure states, $x=\frac{\lambda_2}{\lambda_4}$, $y=\frac{\lambda_3}{\lambda_4}$ with $\lambda_2$, $\lambda_3$, $\lambda_4$ the coefficients given in Eq. (\ref{Ex}).}\label{2}
\end{figure}

The polygamy inequality of entanglement, $C^2_{A|BC}\leq {C_a}^2_{AB}+{C_a}^2_{AC}$, was first introduced in \cite{gour} for a three-qubit pure state $|\psi\rangle_{ABC}$. Since concurrence is equal to the concurrence of assistance for pure states, one gets ${C_a}_{A|BC}\leq {C_a}_{AB}+{C_a}_{AC}$. From Eq. (\ref{m4}) one obtains ${C_a}_{A|BC}\leq (\sqrt{2}-1)\min\{{C_a}_{AB}, {C_a}_{AC}\}+\max\{{C_a}_{AB}, {C_a}_{AC}\}$.
Expectedly, our result is better than ${C_a}_{A|BC}\leq {C_a}_{AB}+{C_a}_{AC}$ derived from Ref. \cite{gour} except for the states such that $\min\{{C_a}_{AB}, {C_a}_{AC}\}=0$.

As another quantum correlation, we consider the tangle of assistance defined by
$\tau_a(\rho_{AB})=\max_{\{p_i,|\psi_i\rangle_{AB}\}}\sum_ip_i\tau(|\psi_i\rangle_{AB})$,
where the tangle $\tau(|\psi\rangle_{AB})$ is given by $\tau(|\psi\rangle_{AB})={2\left[1-\mathrm{Tr}(\rho_A^2)\right]}$ \cite{ckw,lzg}, and the maximum is taken over all possible pure-state decompositions of $\rho_{AB}=\sum\limits_{i}p_i|\psi_i\rangle\langle\psi_i|$ with $p_i\geq0$, $\sum\limits_{i}p_i=1$. Similarly, we have
\begin{eqnarray}\label{Ex2}
\gamma=\frac{\lambda_2^2}{\lambda_2^2+\lambda_4^2}.
\end{eqnarray}
Thus, from Eq. (\ref{Ex2}), we have $\gamma_{\tau_a}=1$ for W-type states that
satisfy the equality in (\ref{Theorem1}) with polygamy power $\beta_{\tau_a}=1$.

From the above calculation, we have $\gamma_{C_a}<\gamma_{\tau_a}$, which implies $\tau_a \preceq C_a$, i.e., the polygamy ability of concurrence of assistance is stronger than that of tangle of assistance. Whereas the polygamy power of the tangle of assistance is smaller than that of concurrence of assistance, i.e., $\beta_{\tau_a}<\beta_{C_a}$, since they are inversely proportional (Fig. 2). Here, we present a definition for polygamy relation of entanglement with equality in Eq. (\ref{m4}), which consists of a substantial and definitive addition to the present understanding of the properties for quantum correlations. It also coincides with the results in \cite{jinzx}: there exist real numbers $\alpha$ and $\beta$ such that the quantum correlation measure $Q^x$ $(x\geq \alpha)$ and $Q^y$ $(0\leq y\leq \beta)$ satisfy monogamy and polygamy relations, respectively.

\section{Polygamy relation of quantum entanglement}

From definition (\ref{m4}) and Theorem 1, generally, an entanglement measure $E$ may not be polygamous (also see \cite{guoy}). Taking the concurrence $C$ as an example, for the state (\ref{Ex}) one obtains $\gamma_C=\sqrt{1+x^2+y^2}-x$, where $x=\frac{\lambda_2}{\lambda_3}$ and $y=\frac{\lambda_4}{\lambda_3}$. When $\lambda_3\to 0$, $\gamma_C \to \infty$, i.e., $C$ cannot be polygamous. In fact, consider separable states with $E_{AC}=0$. Eq. (\ref{m4}) will not hold unless $E_{A|BC}=E_{AB}$ for all separable states. This is not possible for separable states like $|\psi\rangle_{ABC}=\frac{1}{\sqrt{3}}(|000\rangle+|110\rangle+|111\rangle)$, for which we have $C_{A|BC}=\frac{2\sqrt{2}}{3}> C_{AB}\approx0.667$ and $C_{AC}=0$ by using the formula of concurrence for a two-qubit mixed state $\rho$, $C(\rho) = \mathrm{max}\{0,~\eta_1-\eta_2 -\eta_3 -\eta_4\}$, with $\eta_1$, $\eta_2$, $\eta_3$ and $\eta_4$ the square roots of the eigenvalues of $\rho(\sigma_y \otimes \sigma_y )\rho^{\star}(\sigma_y \otimes \sigma_y )$ in nonincreasing order, $\sigma_y$ is the Pauli matrix and $\rho^{\star}$ is the complex conjugate of $\rho$. Another example is the generalized $n$-partite GHZ-class states admitting the multipartite Schmidt decomposition \cite{gjl,ghz}, $|\psi\rangle_{A_1A_2\cdots A_n}=\sum_{i=1}\lambda_i|i_1\rangle\otimes|i_2\rangle\otimes\cdots \otimes|i_n\rangle$, $\sum_i\lambda_i^2=1$, $\lambda_i>0$. These states always give rise to non-polygamous relations for any entanglement measures, since $E(|\psi\rangle_{A_1|A_2\cdots A_n})>0$ and $E(\rho_{A_1A_i})=0$ for all $i=2,\cdots,n$ and any entanglement measure $E$.

Although an entanglement measure $E$ may not be polygamous, we can consider the restrictions among all one-to-group entanglements, the entanglements between a single partite and the remaining ones in an arbitrary multipartite system \cite{qian}. We take tripartite states $\rho_{ABC}$ as an example. Assume $E_{A|BC}\geq E_{B|AC}\geq E_{C|AB}$. $E$ is either polygamous or non-polygamous under the restrictions among all one-to-group entanglements, if $E_{A|BC}\leq E_{B|AC}+E_{C|AB}$ is satisfied or not satisfied. When $E_{C|AB}=0$, one has
$E_{A|BC}= E_{B|AC}$, and $E_{A|BC}\leq E_{B|AC}+E_{C|AB}$ is obviously satisfied. For the case $E_{C|AB}>0$, $E_{A|BC}\leq E_{B|AC}+E_{C|AB}$ may not hold for some entanglement measures. In the following, we introduce a definition for polygamy relation with equality among all one-to-group entanglements.

{\bf Definition 2.} Let $E$ be a measure of quantum entanglement. $E$ is said to be polygamous if for any state $\rho_{ABC}$,
\begin{eqnarray}\label{poly}
E_{A|BC}= E_{B|AC}+\delta\, E_{C|AB}
\end{eqnarray}
for some $\delta>0$, where $E_{A|BC}\geq E_{B|AC}\geq E_{C|AB}$. We call $\delta$ the polygamy weight with respect to the entanglement measure $E$.

From the definition, if $E_{C|AB}=0$ for any quantum state $\rho_{ABC}$, Eq. (\ref{poly}) holds for any $\delta>0$. Otherwise, one can always choose $\delta(\rho_{ABC})=\frac{E_{A|BC}-E_{B|AC}}{E_{C|AB}}$ such that entanglement measure $E$ is polygamous. Thus, there always exists a $\delta$ ($\delta=\max_{\rho_{ABC}}\delta(\rho_{ABC})$) such that $E_{A|BC}\leq E_{B|AC}+\delta E_{C|AB}$ is satisfied for all states $\rho_{ABC}$.

In Ref. \cite{qian}, the authors investigate the polygon relationship among the one-to-group marginal entanglement measures of pure qubit systems. Entanglement measures, including von Neumann entropy [16], concurrence [17], negativity [18], and the normalized Schmidt weight satisfy the symmetric inequalities $E_{i|jk}\leq E_{j|ik}+E_{k|ij}$ with $i\ne j\ne k \in \{A,B,C\}$. In fact, the result in \cite{qian} is a special case of Eq. (\ref{poly}) for $\delta=1$. The cases with $\delta<1$ are beyond the results in \cite{qian}. Moreover, Eq. (\ref{poly}) not only applies to pure states but also to mixed states.
In the following, we use concurrence as an example to show the advantage of Eq. (\ref{poly}).

Let us consider the state $|\psi\rangle_{ABC}=\sin\theta\cos\phi|000\rangle+\sin\theta\sin\phi|101\rangle +\cos\theta|110\rangle$. We have $C_{A|BC}=2\sin\theta\cos\phi\sqrt{\sin^2\theta\sin^2\phi+\cos^2\theta}$, $C_{B|AC}=\sin2\theta$ and $C_{C|AB}=2\sin\theta\sin\phi\sqrt{\sin^2\theta\cos^2\phi+\cos^2\theta}$. From Eq. (\ref{poly}), we obtain $\delta_C=\frac{\cos\phi\sqrt{\sin^2\theta\sin^2\phi
+\cos^2\theta}-\cos\theta}{\sin\phi\sqrt{\sin^2\theta\cos^2\phi+\cos^2\theta}}$.
Suppose $C_{A|BC}\geq C_{B|AC}\geq C_{C|AB}$, which implies that the parameters $\theta$ and $\phi$ are in the triangle-like region shown in Fig. 4. If $|\psi\rangle_{ABC}$ is a separable state, it corresponds to the point $P$ in Fig. 4, that is, either (i) $\phi=0$ or (ii) $\theta=\frac{\pi}{2}$ or (iii) $\theta=\frac{\pi}{2}$ and $\phi=0$. For example, when $\phi=0$, the related state $|\psi\rangle_{ABC}$ is changed into a separable state $\sin\theta|000\rangle +\cos\theta|110\rangle$. For all these cases, the polygamy weight $\delta_C=1$; otherwise $\delta_C<1$. For example, setting $\theta=\phi=\frac{\pi}{4}$ one obtains $\delta_C=\frac{2}{\sqrt{3}}-1<1$.
\begin{figure}
\centering
\includegraphics[width=8.5cm]{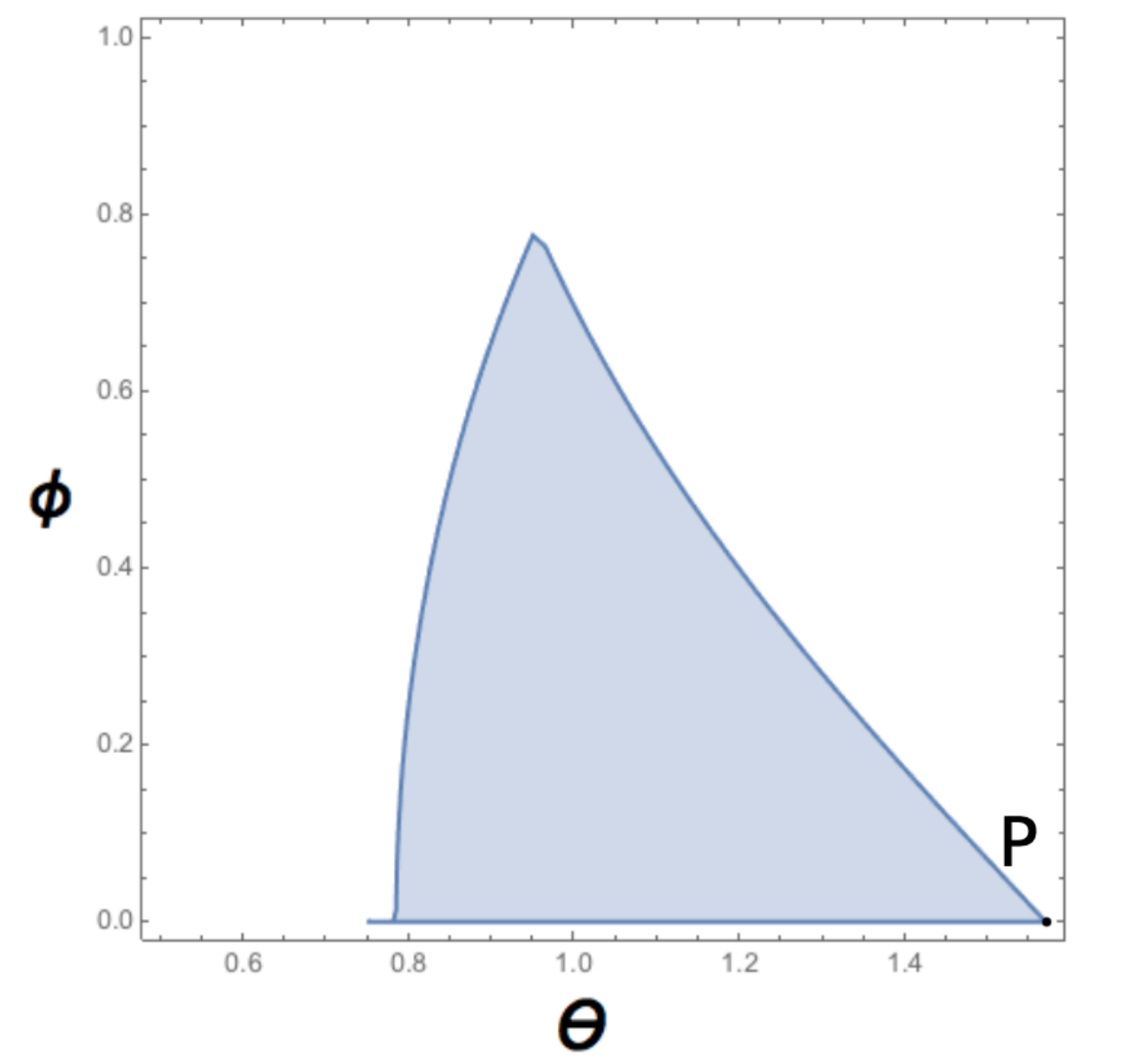}\\
\caption{Triangle-like region for the parameters $\theta$ and $\phi$ such that the inequality $C_{A|BC}\geq C_{B|AC}\geq C_{C|AB}$ is satisfied.}\label{2}
\end{figure}

Remark: In the above example, we only discussed the condition $C_{A|BC}\geq C_{B|AC}\geq C_{C|AB}$. In fact, there are other five cases, $C_{A|BC}\geq C_{C|AB}\geq C_{B|AC}$, $C_{B|AC}\geq C_{A|BC}\geq C_{C|AB}$, $C_{B|AC}\geq C_{C|AB}\geq C_{A|BC}$, $C_{C|AB}\geq C_{A|BC}\geq C_{B|AC}$, and $C_{C|AB}\geq C_{B|AC}\geq C_{A|BC}$. One can similarly analyze these cases and obtain the same results.

Concerning the polygamy power $\eta$, we have the following familiar inequality (see the proof in Appendix).

{\bf Theorem 3}  Let $E$ be a measure of entanglement. $E$ is polygamous according to the definition (\ref{poly}) if and only if there exits $\eta >0$ such that
\begin{eqnarray}\label{th2}
E^\eta_{A|BC}\leq E^\eta_{B|AC}+ E^\eta_{C|AB}
\end{eqnarray}
for any state $\rho_{ABC}$.

The polygamy weight $\gamma$ functions as a bridge in characterizing the polygamous ability among different quantum correlation measures. A quantum correlation measure $Q$ with a smaller $\gamma$ is more likely to be polygamous. Thus, the polygamy weight $\gamma$ gives the physical meaning of the coefficients introduced in Ref. \cite{jzx1} for the weighted polygamy relations. Particularly, quantum entanglement measures cannot be polygamous since the entanglement of the reduced GHZ-class states is 0. Nevertheless, we provide polygamy relations with equality and obtain a polygamy relation for tripartite systems based on one-to-group entanglement.

\section{Conclusion}

Quantum correlation is a fundamental property of multipartite systems. For a given quantum correlation measure $Q$, we have introduced a new definition of the polygamy relation with equality, which characterizes the precise division of the quantum correlation distribution. The non-polygamous quantum correlation distribution is only located on the coordinate axis, as shown in Fig. 1; the blue region satisfies both our notion of polygamy (\ref{m4}) and the usual one (\ref{m1}), whereas the orange, yellow, and white regions violate the inequality (\ref{m1}) but still satisfy our polygamy relation (\ref{m4}). The advantage of our notion of polygamy is that one can determine which quantum correlation measure is more likely to be polygamous by comparing the polygamy weights. We have used the concurrence of assistance and tangle of assistance as examples, showing that the concurrence of assistance is more likely to be polygamous than the tangle of assistance since the weight of the tangle of assistance is larger than that of concurrence of assistance. However, by using $Q^\beta$ for some $\beta>0$, we have shown that our polygamy relation can reproduce the conventional polygamy inequalities such as (\ref{m1}). Furthermore, it has been shown that quantum entanglement measures cannot be polygamous as the GHZ-class reduced states are separable. Inspired by Ref. \cite{qian}, considering the one-to-group entanglements between a single partite and the remaining ones, we have provided the polygamy relations with equality and obtained a polygamy relation for tripartite systems. Generalizing our results to multipartite systems, we have obtained Theorem 2 for $N$-partite states. Our results may shed new light on polygamy properties related to other quantum correlations.

\bigskip
\noindent{\bf Acknowledgments}\, \,
This work was supported in part by the National Natural Science Foundation of China (NSFC) under Grants 12301582, 12075159, 12171044, 11975236, and 12235008; Beijing Natural Science Foundation (Grant No. Z190005); Academician Innovation Platform of Hainan Province; Guangdong Basic and Applied Basic Research Foundation 2020A1515111007; Start-up Funding of Guangdong Polytechnic Normal University No. 2021SDKYA178; Start-up Funding of Dongguan University of Technology No. 221110084.

\bigskip
\section*{APPENDIX}
\setcounter{equation}{0} \renewcommand%
\theequation{A\arabic{equation}}

\subsection{Proof of Theorem 1}

Let $Q$ be a polygamous measure of the quantum correlation that satisfies (\ref{m4}). If $Q_{A|BC}=0$ or $Q_{A|BC}= Q_{AC}$, the result is obvious. Assume $Q_{A|BC}> 0$ and $Q_{A|BC}> Q_{AC}$. Since $Q$ is a measure of quantum correlation, it is non-increasing under a partial trace. Then, we have $Q_{AC}\geq Q_{AB}>0$ for $\gamma\geq0$ according to (\ref{m4}). Thus, $Q_{A|BC}> Q_{AC}\geq Q_{AB}>0$ for any state $\rho_{ABC}$.

Set $g(\beta)=Q^\beta_{A|BC}-Q^\beta_{AB}-Q^\beta_{AC}$ with $0< \beta \leq 1$. From Eq. (\ref{m4}), we have
\begin{eqnarray}\label{pfth11}
g(\beta)&=&(\gamma Q_{AB}+Q_{AC})^\beta-Q^\beta_{AB}-Q^\beta_{AC}\nonumber\\
&=&Q^\beta_{AC}(1+\gamma\frac{Q_{AB}}{Q_{AC}})^\beta-Q^\beta_{AB}-Q^\beta_{AC}\nonumber\\
&\leq&Q^\beta_{AC}[1+\beta(\gamma\frac{Q_{AB}}{Q_{AC}})^\beta]-Q^\beta_{AB}-Q^\beta_{AC}\nonumber\\
&=&(\beta\gamma^\beta-1)Q^\beta_{AB},
\end{eqnarray}
where the inequality is due to $(1+t)^x\leq 1+xt^x$ \cite{finer} with $0\leq x,t\leq 1$ and $0\leq \gamma\leq \frac{Q_{AC}}{Q_{AB}}$.

Now, we only need to prove that there exists a $\beta$ such that $\beta\gamma^\beta\leq1$. From the definiton (\ref{m4}), the polygamy inequality (\ref{m1}) is obviously true for $0\leq \gamma \leq 1$. For $1<\gamma \leq \frac{Q_{AC}}{Q_{AB}}$, one can always choose $\beta=\frac{Q_{AB}}{Q_{AC}+1}$ such that $\beta\gamma^\beta<1$ (Fig. 5). Thus, the inequality is proved (\ref{Theorem1}).
\begin{figure}
\centering
\includegraphics[width=8.5cm]{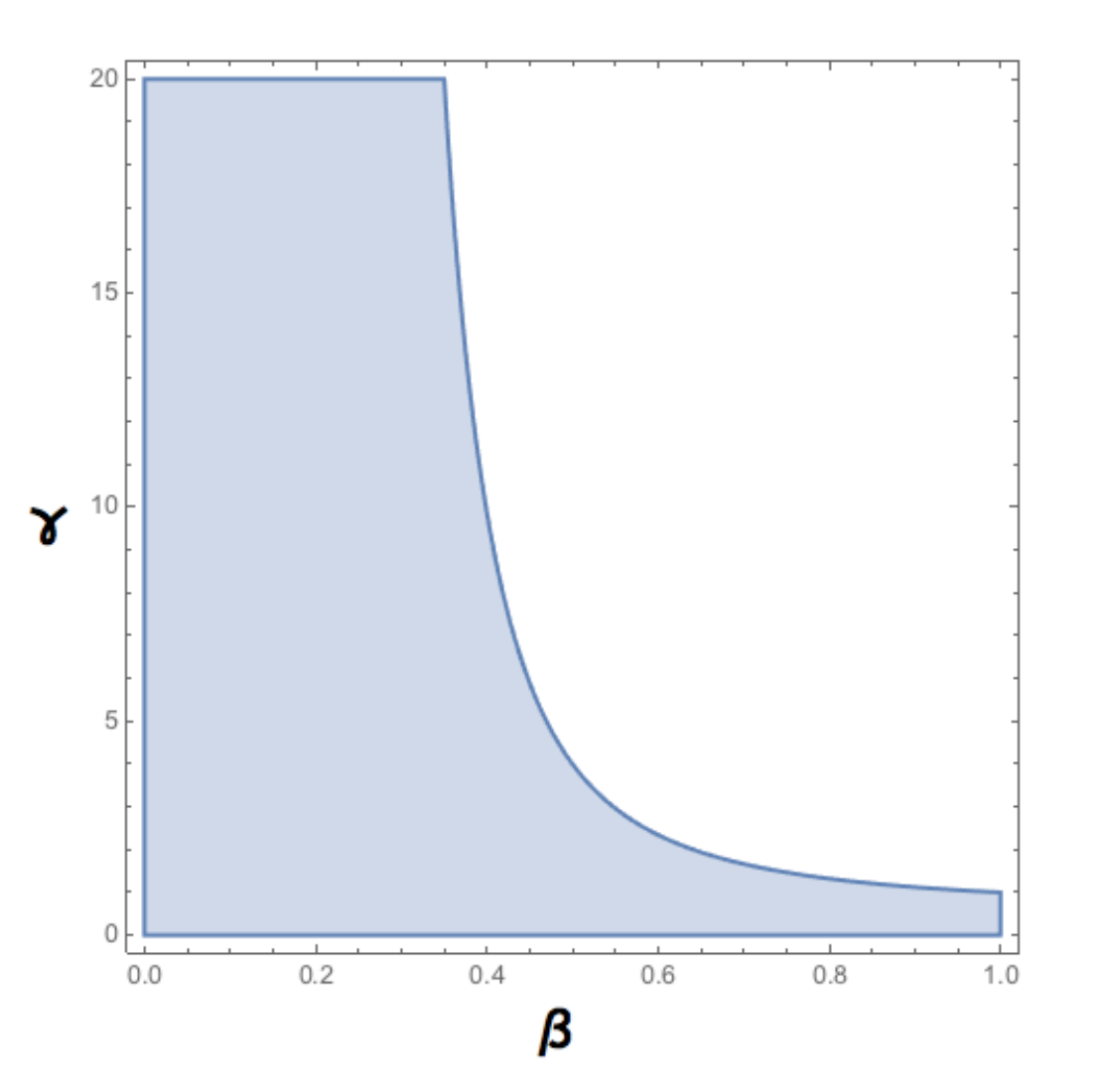}\\
\caption{Blue region denotes the point ($\beta$, $\gamma$) satisfying the inequality $\beta\gamma^\beta\leq1$.}\label{2}
\end{figure}

Now assume $Q^\beta_{A|BC}\leq Q^\beta_{AB}+Q^\beta_{AC}$ ($\beta>0$) for any state $\rho_{ABC}$ and $Q_{AC}\geq Q_{AB}$. Since $Q$ is a quantum correlation measure, we have $Q_{A|BC}\geq Q_{AC}\geq Q_{AB}$. If $Q_{AB}=0$, then $Q_{A|BC}= Q_{AC}$ due to the inequality (\ref{Theorem1}). If $Q_{AB}>0$, then $Q_{A|BC}\geq Q_{AC}\geq Q_{AB}>0$. We can always choose $\gamma=\max_{\rho_{ABC}}\frac{Q_{A|BC}- Q_{AC}}{Q_{AB}}$ such that Eq. (\ref{m4}) holds. This completes the proof.

\bigskip

\subsection{Proof of Theorem 2}

From (\ref{m4}), we have
\begin{eqnarray}\label{pf41}
&&Q_{A|B_1B_2\cdots B_{N-1}}\nonumber\\
&&= Q_{AB_1}+\gamma_1Q_{A|B_2\cdots B_{N-1}}\nonumber\\
&&= Q_{AB_1}+\gamma_1 Q_{AB_2}+\gamma_1\gamma_2Q_{A|B_3\cdots B_{N-1}}\nonumber\\
&&= \cdots\nonumber\\
&&= Q_{AB_1}+\gamma_1 Q_{AB_2}+\cdots+\gamma_1\gamma_2\cdots \gamma_{m-1}Q_{AB_m}\nonumber\\
&&~~~+\gamma_1\gamma_2\cdots \gamma_mQ_{A|B_{m+1}\cdots B_{N-1}}.
\end{eqnarray}
Similarly, because ${Q_{AB_j}}\leq {Q_{A|B_{j+1}\cdots B_{N-1}}}$ for $j=m+1,\cdots,N-2$, we obtain
\begin{eqnarray}\label{pf42}
&&Q_{A|B_{m+1}\cdots B_{N-1}} \nonumber\\
&&= \gamma_{m+1}Q_{AB_{m+1}}+Q_{A|B_{m+2}\cdots B_{N-1}}\nonumber\\
&&= \gamma_{m+1}Q_{AB_{m+1}}+\cdots+\gamma_{N-2}Q_{AB_{N-2}}\nonumber\\
&&~~~+Q_{AB_{N-1}}.
\end{eqnarray}
Combining (\ref{pf41}) and (\ref{pf42}), we obtain Theorem 2.

\subsection{Proof of Theorem 3}

Let $E$ be a polygamous measure of quantum entanglement that satisfies (\ref{poly}). If $E_{X|\bar{X}}=0$ for $X=A,B,C$, the result is obvious. Assume $E_{A|BC}> 0$ and $E_{A|BC}\geq E_{B|AC}\geq E_{C|AB}$. Set $g(\eta)=E^\eta_{A|BC}-E^\eta_{B|AC}-E^\eta_{C|AB}$, with $0< \eta \leq 1$. From Eq. (\ref{poly}), we have
\begin{eqnarray}\label{pfth11}
g(\eta)&=&(E_{B|AC}+\delta E_{C|AB})^\eta-E^\eta_{B|AC}-E^\eta_{C|AB}\nonumber\\
&=&E^\eta_{B|AC}(1+\delta\frac{E_{C|AB}}{E_{B|AC}})^\eta-E^\eta_{B|AC}-E^\eta_{C|AB}\nonumber\\
&\leq&E^\eta_{B|AC}[1+\eta(\delta\frac{E_{C|AB}}{E_{B|AC}})^\eta]-E^\eta_{B|AC}-E^\eta_{C|AB}\nonumber\\
&=&(\eta\delta^\eta-1)E^\eta_{C|AB},
\end{eqnarray}
where the inequality is due to $(1+t)^x\leq 1+xt^x$ for $0\leq x,t\leq 1$ and $0\leq \delta \leq \frac{E_{B|AC}}{E_{C|AB}}$.

Now, we only need to prove that there exists an $\eta$ such that $\eta\delta^\eta\leq1$. From the definition (\ref{poly}), the polygamy inequality (\ref{m1}) is obviously true for $0\leq \delta \leq 1$. For $1<\delta \leq \frac{E_{B|AC}}{E_{C|AB}}$, one can always choose $\eta=\frac{E_{C|AB}}{E_{B|AC}+1}$ such that $\eta\delta^\eta<1$. Thus, the inequality is proved (\ref{th2}).

Assume $E^\eta_{A|BC}\leq E^\eta_{B|AC}+ E^\eta_{C|AB}$ ($\eta>0$) for any state $\rho_{ABC}$ and $E_{A|BC}\geq E_{B|AC}\geq E_{C|AB}$. From Eq. (\ref{poly}), one has $E_{A|BC}= E_{B|AC}$ if $E_{C|AB}=0$. If $E_{C|AB}>0$, one may choose $\delta=\max_{\rho_{ABC}}\frac{E_{A|BC}- E_{B|AC}}{E_{C|AB}}$ such that Eq. (\ref{poly}) holds. The proof is complete.

\end{document}